\documentclass[twocolumn, superscriptaddress, floatfix, showpacs, aps, prb, 10pt]{revtex4-2}
\pdfoutput=1
\usepackage{graphicx}
\usepackage{amsmath}
\usepackage[colorlinks=true, citecolor=blue, urlcolor=blue]{hyperref}
\usepackage{amssymb}
\usepackage{bm}
\usepackage{color}
\usepackage{array}
\usepackage{booktabs}
\usepackage{multirow}
\usepackage{braket}

\newcommand{\pd}{{\phantom{\dag}}}

\begin{document}

\title{Nontrivial gapless electronic states at the stacking faults of weak topological insulators}

\author{Gabriele Naselli}
\affiliation{Leibniz Institute for Solid State and Materials Research (IFW) Dresden, Helmholtzstrasse 20, 01069 Dresden, Germany}
\affiliation{W{\"u}rzburg-Dresden Cluster of Excellence ct.qmat, Helmholtzstrasse 20, 01069 Dresden, Germany}

\author{Viktor K\"{o}nye}
\affiliation{Leibniz Institute for Solid State and Materials Research (IFW) Dresden, Helmholtzstrasse 20, 01069 Dresden, Germany}
\affiliation{W{\"u}rzburg-Dresden Cluster of Excellence ct.qmat, Helmholtzstrasse 20, 01069 Dresden, Germany}

\author{Sanjib Kumar Das}
\affiliation{Leibniz Institute for Solid State and Materials Research (IFW) Dresden, Helmholtzstrasse 20, 01069 Dresden, Germany}
\affiliation{Department of Physics, Lehigh University, Bethlehem, Pennsylvania, 18015, USA}

\author{G. G. N. Angilella}
\affiliation{Dipartimento di Fisica e Astronomia ``Ettore Majorana'', Universit\`a di Catania, 64, Via S. Sofia, I-95123 Catania, Italy}
\affiliation{Scuola Superiore di Catania, Universit\`a di Catania, 9, Via Valdisavoia, I-95123 Catania, Italy}
\affiliation{INFN, Sez. Catania, 64, Via S. Sofia, I-95123 Catania, Italy}

\author{Anna Isaeva}
\affiliation{Leibniz Institute for Solid State and Materials Research (IFW) Dresden, Helmholtzstrasse 20, 01069 Dresden, Germany}
\affiliation{Van der Waals-Zeeman Institute, Department of Physics and Astronomy, University of Amsterdam, Science Park 094, 1098 XH Amsterdam, The Netherlands}

\author{Jeroen van den Brink}
\affiliation{Leibniz Institute for Solid State and Materials Research (IFW) Dresden, Helmholtzstrasse 20, 01069 Dresden, Germany}
\affiliation{W{\"u}rzburg-Dresden Cluster of Excellence ct.qmat, Helmholtzstrasse 20, 01069 Dresden, Germany}

\author{Cosma Fulga}
\affiliation{Leibniz Institute for Solid State and Materials Research (IFW) Dresden, Helmholtzstrasse 20, 01069 Dresden, Germany}
\affiliation{W{\"u}rzburg-Dresden Cluster of Excellence ct.qmat, Helmholtzstrasse 20, 01069 Dresden, Germany}

\begin{abstract}
Lattice defects such as stacking faults may obscure electronic topological features of real materials. 
In fact, defects are a source of disorder that can enhance the density of states and conductivity of the bulk of the system and they break crystal symmetries that can protect the topological states. 
On the other hand, in recent years it has been shown that lattice defects can act as a source of nontrivial topology.
Motivated by recent experiments on three-dimensional (3D) topological systems such as Bi$_2$TeI and Bi$_{14}$Rh$_3$I$_9$, we examine the effect of stacking faults on the electronic properties of weak topological insulators (WTIs).
Working with a simple model consisting of a 3D WTI formed by weakly-coupled two-dimensional (2D) topological layers separated by trivial spacers, we find that 2D stacking faults can carry their own, topologically nontrivial gapless states.
Depending on the WTI properties, as well as the way in which the stacking fault is realized, the latter can form a topologically protected 2D semimetal, but also a 2D topological insulator which is embedded in the higher-dimensional WTI bulk.
This suggests the possibility of using stacking faults in real materials as a source of topologically nontrivial, symmetry-protected conducting states.
\end{abstract}

\maketitle

\section{Introduction}
\label{sec:intro}

Topological phases hosting the quantum Hall effect \cite{vonKlitzing1980, Thouless1982} or the quantum spin-Hall effect (QSHE) \cite{Kane2005, Kane2005_2, Bernevig2006, Konig2007} are characterized by an insulating bulk and gapless, topologically-protected boundary states.
It is possible to classify different topological phases depending on the symmetries protecting these boundary states and the topology of the bulk Hamiltonian \cite{Kitaev2009, Ryu2010}. 
In a so-called strong topological insulator, only local symmetries (chiral, time-reversal, or particle-hole symmetry) are required to protect the edge states \cite{Hasan2010}.
In recent years, it was found out that some types of topological insulators, such as weak topological insulators (WTIs) \cite{Fu2007, Moore2007, Roy2009}, higher-order topological insulators \cite{Benalcazar2017, Schindler2018}, and topological crystalline insulators \cite{Fu2011, Hsieh2012, Ando2015} have boundary states protected by lattice symmetries \cite{Zhang2018}, in addition to the local ones.

The presence of defects in topological insulators can be detrimental to the experimental detection of their topological properties \cite{Fulga2014}.
Point-like impurities, for instance, can increase both the the density of states as well as the conductivity of the otherwise insulating bulk of the material. 
This can make it difficult to separate bulk and boundary contributions in transport.
Furthermore, defects break lattice symmetries, which means they may be doubly detrimental for the experimental detection of the topological properties of materials which rely on these lattice symmetries to protect the conducting states.

In spite of this, lattice defects can carry topologically nontrivial states in and of themselves \cite{Teo2010}. 
This has been shown to occur for a variety of so-called `topological defects', such as full dislocations \cite{Teo2010, Ran2009, Teo2017, Jurii2012, Hughes2014} and disclinations \cite{Teo2013, Benalcazar2014}, which break lattice symmetries only locally and for this reason are locally undetectable far from their core.
The resulting classification of topological defects now includes those present in strong and weak topological insulators \cite{Ringel2012, Tretiakov2010, Sbierski2016, Hamasaki2017}, as well as in topological crystalline phases \cite{Benalcazar2014, Teo2013, Shiozaki2014, Slager2014} and higher-order topological insulators \cite{Queiroz2019, Roy2021}.

In contrast to topological defects, which are by now well understood, nontopological defects, which produce large-scale, visible distortions of the lattice, have only recently begun drawing attention. 
Partial dislocations, which lead to the formation of stacking faults, have been shown to host topologically protected modes in higher-order topological insulators \cite{Queiroz2019, Yamada2021}.
Even when the bulk of the material itself is trivial, it has been shown that stacking faults can form lower-dimensional topologically nontrivial subsystems.
In such cases, the stacking faults present in the otherwise trivial bulk have been dubbed `embedded' \cite{Tuegel2019, Velury2021} topological insulators and semimetals.

In this work, we focus instead on weak topological insulators, which have recently been experimentally realized in the Bi$_2$TeI \cite{Rusinov2016} and Bi$_{14}$Rh$_3$I$_9$ \cite{Rasche2013} material families. 
In Bi$_{14}$Rh$_3$I$_9$, the nontrivial layer is a ([Bi$_4$Rh]$_3$I)$^{2+}$ honeycomb lattice, which is topologically equivalent to graphene \cite{Rasche2016}, and the spacer is formed by two coupled, one-dimensional (BiI$_4$)$^{-}$ chains. 
There is one spacer fragment per honeycomb QSHE entity. In Bi$_2$TeI, the QSHE is hosted by bismuth bilayers Bi$_2$ \cite{Murakami2006, Avraham2020}. 
If interlayer coupling is disregarded in this system, the QSHE bismuth bilayers are separated by two consecutive BiTeI spacer layers that are topologically trivial. 
In the real Bi$_2$TeI material, a topological band inversion is realized via interlayer interaction between the Bi$_2$ and BiTeI layers that results in the formation of QSHE sandwiches BiTeI $\cdot$ Bi$_2$ $\cdot$ BiTeI \cite{Rusinov2016}. 
This coupling is further strengthened when the Bi$_2$-to-BiTeI layer ratio reaches 1:1 in Bi$_3$TeI which shows an electronic spectrum of a topological metal \cite{Zeugner2017}. 
Interestingly, the chemical composition (bromine for iodine substitution) and the number of spacer layers (one or two) per Bi$_2$ QSHE layer can be controlled by crystal growth parameters~\cite{Zeugner2017, Zeugner2018}. 
Here, we model these systems as weakly-coupled stackings of QSHE layers and a varying number of trivial spacers.

Given the recent advances in the growth and control of tailor-made topological materials, we examine the phenomenology of stacking faults in WTIs from a theoretical point of view.
These defects break the lattice translation symmetry, which is required to protect the WTI phase.
Nevertheless, we show that stacking faults can carry their own topologically nontrivial modes, forming either two-dimensional (2D) semimetal phases or 2D QSHE phases which are embedded within the 3D bulk of the WTI.
Which type of topology is realized depends on the properties of the stacking fault itself, namely on the fractional lattice translation used to realize it, and it also depends on the number of spacer layers in the parent WTI.

In what follows, we start by introducing a tight-binding model to describe the topological properties of the WTIs without a stacking fault (Sec.~\ref{sec:model}). We explain then how to construct a stacking fault in our model and in which cases it is possible to have topological states at the stacking fault plane using a heuristic argument (Sec.~\ref{sec:stacking}). In Sec.~\ref{sec:numerics}, we solve numerically the tight-binding model in the presence of a stacking fault in WTIs with different numbers of spacer layers, showing that the numerical results support our initial argument and that it is possible to obtain embedded 2D semimetals and topological insulators at the defect. 
We conclude and discuss directions for future work in Sec.~\ref{sec:conc}.

\section{Weak topological insulator model}
\label{sec:model}

Throughout this paper, we will work with a WTI tight-binding system consisting of a 3D stack of weakly-coupled, 2D QSHE layers separated by trivial spacers.
Each of the 2D layers is a honeycomb lattice, with sublattices labeled $A$ and $B$, and the stacking is of the $A-A$ type. We choose the lattice constant such as to obtain Bravais vectors ${\bf a_x}=(1, 0, 0)$, ${\bf a_y}=(1/2, \sqrt{3}/2, 0)$, and ${\bf a_z}=(0, 0, 1)$, the latter vector corresponding to the stacking direction. Thus, unit cells are labeled by position vectors 
${\bf r} = n_x {\bf a_x} + n_y {\bf a_y} + n_z {\bf a_z}$, $n_{x,y,z} \in {\mathbb Z}$.

To model the 2D topological layers, we use the well-known QSHE phase realized by the Kane-Mele model \cite{Kane2005, Kane2005_2}. 
The real-space Hamiltonian of each 2D layer (thus at constant $n_z$) reads
\begin{equation}
\begin{split}
H_{\rm KM} & = \sum_ {\rm N.N.} t\, c^\dagger_{{\bf r}, i,\alpha}c^\pd_{{\bf r'}, j,\beta}s^0_{\alpha,\beta} \\
 & +\sum_ {\rm N.N.N.} it^\pd_2 \nu^\pd_{{\bf r}, {\bf r'}, j}c^\dagger_{{\bf r}, j,\alpha}c^\pd_{{\bf r'}, j,\beta}s^z_{\alpha,\beta},
\end{split}
\label{eq:KMmodel}
\end{equation}
where $c^\dagger_{{\bf r}, j,\alpha}$ creates an electron with spin $\alpha=\uparrow, \downarrow$ on the $j=A,B$ sublattice of the unit cell at position ${\bf r}$. 
The Pauli matrices $s^{0,x,y,z}$ encode the spin degree of freedom, whereas N.N. and N.N.N. denote nearest and next-nearest neighbor sites \textit{in the plane} of the honeycomb lattice. 
In the Kane-Mele model, the next-nearest-neighbor term $t_2$ opens a gap in the graphene-like spectrum of the model, leading to a topological insulating phase \cite{Kane2005, Kane2005_2}. 
The coefficient $\nu^\pd_{{\bf r}, {\bf r'}, j}$ is either $+1$ or $-1$ depending on whether hoppings go in the clockwise or counter-clockwise direction along a hexagonal plaquette of the honeycomb lattice. 

To model the trivial spacers in the WTI, we use a honeycomb lattice with only nearest-neighbor hoppings and add an on-site term which has an opposite sign for sites belonging to different sublattices. 
Similar to the Kane-Mele term, this gaps out the Dirac cones at the corners of the Brillouin zone \cite{Kane2005, Kane2005_2, Goerbig2011}, but it produces a trivial system.
The 2D Hamiltonian of each spacer, again at constant $n_z$, is
\begin{equation}
H_{\rm I}=\sum_ {\rm N.N.} t\, c^\dagger_{{\bf r}, i,\alpha} c^\pd_{{\bf r'}, j,\beta} s^0_{\alpha,\beta} + \sum_ {{\bf r}, j, \alpha} m \lambda^\pd_j c^\dagger_{{\bf r}, j,\alpha} c^\pd_{{\bf r}, j,\alpha},
\label{eq:insham}
\end{equation}
where $m$ is a constant, $\lambda^\pd_j=\pm 1$ for the $A$ and $B$ sublattice, respectively, and the first sum runs, as before, over the nearest-neighbor sites within a given layer.

Using the two ingredients above, we can now construct a tight binding model for a WTI whose unit cell is composed of a single QSHE layer and a single spacer \cite{Das2019}.
Thus, the in-plane Hamiltonian takes the form of Eq.~\eqref{eq:KMmodel} whenever $n_z$ is even, and the form of Eq.~\eqref{eq:insham} whenever $n_z$ is odd.
We add spin-dependent hoppings in the ${\bf a_z}$ direction between the layers, in such a way that the edge states moving in opposite directions in adjacent topological layers will couple with each other. 
The inter-layer coupling term takes the form
\begin{equation}
\begin{split}
H_z = & \sum_{{\bf r}, j, \alpha, \beta} it^\pd_z c^\dagger_{{\bf r}, j,\alpha} c^\pd_{{\bf r}+{\bf a_z}, j,\beta} s^z_{\alpha, \beta} \\
& + it^\pd_{z2} c^\dagger_{{\bf r}, j,\alpha} c^\pd_{{\bf r}+2{\bf a_z}, j,\beta} s^y_{\alpha, \beta}+ {\rm h. c.}
\end{split}
\label{eq:bihz}
\end{equation}
Thus, the nearest-neighbor hoppings in the ${\bf a_z}$ direction couple the QSHE layers to the neighboring spacers with a strength $t^\pd_z$. We choose a diagonal term in the spin degree of freedom, $s^z$, similar to the one in the Kane-Mele model. 
\begin{figure}[tb]
\centering
\includegraphics[width=8.6cm]{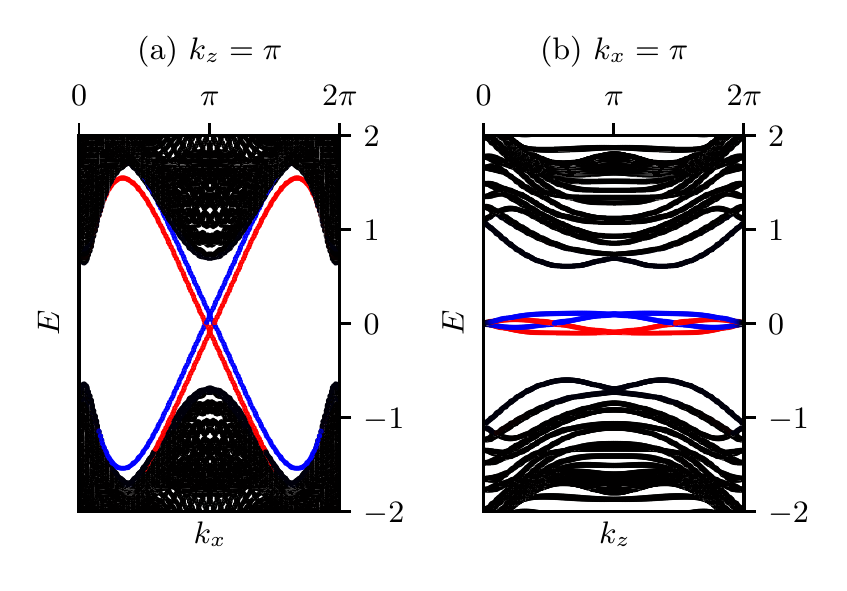}
\caption{
Band structure of the WTI at $k_z=\pi$ (panel a) and at $k_x=\pi$ (panel b).
We use a slab geometry, infinite along ${\bf a_{x}}$ and ${\bf a_{z}}$, but consisting of $L_y=30$ unit cells in the ${\bf a_{y}}$ direction. 
The Hamiltonian parameters are $t_2=1$, $m=2$, $t_z=0.5$, and $t_{z2}=0.2$ in units of $t$. 
Bulk states are shown in black, whereas states localized at opposite surfaces of the slab are shown in red and blue.
}
\label{fig:plot_spectra_twolayer}
\end{figure}
The second-neighbor hopping with strength $t^\pd_{z2} \leq t^\pd_z$, on the other hand, couples the QSHE edge modes in adjacent unit cells, leading to a larger dispersion of the resulting WTI surface states.
The second-neighbor hopping contains a term which is off-diagonal in spin space, $s^y$, since in the Kane-Mele model counter-propagating modes have opposite spin.

We consider an infinite slab geometry, in which the WTI is infinite along the ${\bf a_x}$ and ${\bf a_z}$ direction, but contains a finite number $L_y$ of unit cells in the ${\bf a_y}$ direction. 
We set the nearest-neighbor hopping to $t=1$ and express all other energy scales relative to it. Labeling the dimensionless momentum parallel to the layers as $k_x$ and the one in the stacking direction as $k_z$, we plot the band structure of the system in Fig.~\ref{fig:plot_spectra_twolayer}. 
All of our numerical results are obtained using the Kwant code \cite{Groth2014}, and our code is available on Zenodo at \cite{code}.

As expected, two Dirac cones form on each of the two surfaces of the WTI slab, at $(k_x, k_z)=(\pi, 0)$ and $(\pi, \pi)$, respectively. They connect the valence and conduction bands, as visible in Fig.~\ref{fig:plot_spectra_twolayer}(a).
Notice, however, that at momenta $(\pi, \pi)$ the Dirac points on opposite surfaces (shown in red and blue) have different energies.
This is a consequence of the orientation of the WTI surface, as well as of the staggered on-site potential used to model the spacer layers in Eq.~\eqref{eq:insham}.
Since the slab is chosen to be finite in the ${\bf a_y}$ direction, each of the 2D layers composing it has a zigzag termination.
Thus, on one surface the sites of the spacers belong to the $A$ sublattice and experience a positive on-site potential, whereas on the opposite surface all sites belong to the $B$ sublattice and experience a negative on-site potential.

Finally, based on the same construction as above, we also consider a WTI with a unit cell composed of one topological layer and two spacer layers.
Their Hamiltonians are, as before, given by Eqs.~\eqref{eq:KMmodel} and \eqref{eq:insham}, with the difference that we use the Kane-Mele model whenever $n_z$ is divisible by three and use the trivial Hamiltonian for all other values of $n_z$.
To ensure a good coupling between the edge modes of QSHE layers in adjacent unit cells, we modify the inter-layer coupling Hamiltonian of Eq.~\eqref{eq:bihz} by adding a third-nearest neighbor hopping.
The latter has the same spin structure, $s^y$, and a magnitude $t_{z3} \leq t_{z2}$.
The resulting, inter-layer coupling Hamiltonian is:
\begin{equation}
\begin{split}
\widetilde{H}_z = & \sum_{{\bf r}, j, \alpha, \beta} it^\pd_z c^\dagger_{{\bf r}, j,\alpha} c^\pd_{{\bf r}+{\bf a_z}, j,\beta} s^z_{\alpha, \beta} \\
& + it^\pd_{z2} c^\dagger_{{\bf r}, j,\alpha} c^\pd_{{\bf r}+2{\bf a_z}, j,\beta} s^z_{\alpha, \beta} \\
&+it^\pd_{z3} c^\dagger_{{\bf r}, j,\alpha} c^\pd_{{\bf r}+3{\bf a_z}, j,\beta} s^y_{\alpha, \beta}+ {\rm h. c.}
\end{split}
\label{eq:threehz}
\end{equation}

For the two-spacer WTI, the band structure in a slab geometry (not shown) is qualitatively similar to that of Fig.~\ref{fig:plot_spectra_twolayer}.
Two Dirac cones form on each surface, at the same points of the surface Brillouin zone, and are displaced in energy by the staggered on-site potentials of the spacers.

\section{Stacking fault construction and phenomenology}
\label{sec:stacking}

\begin{figure}[tb]
\centering
\includegraphics[width=8.6cm]{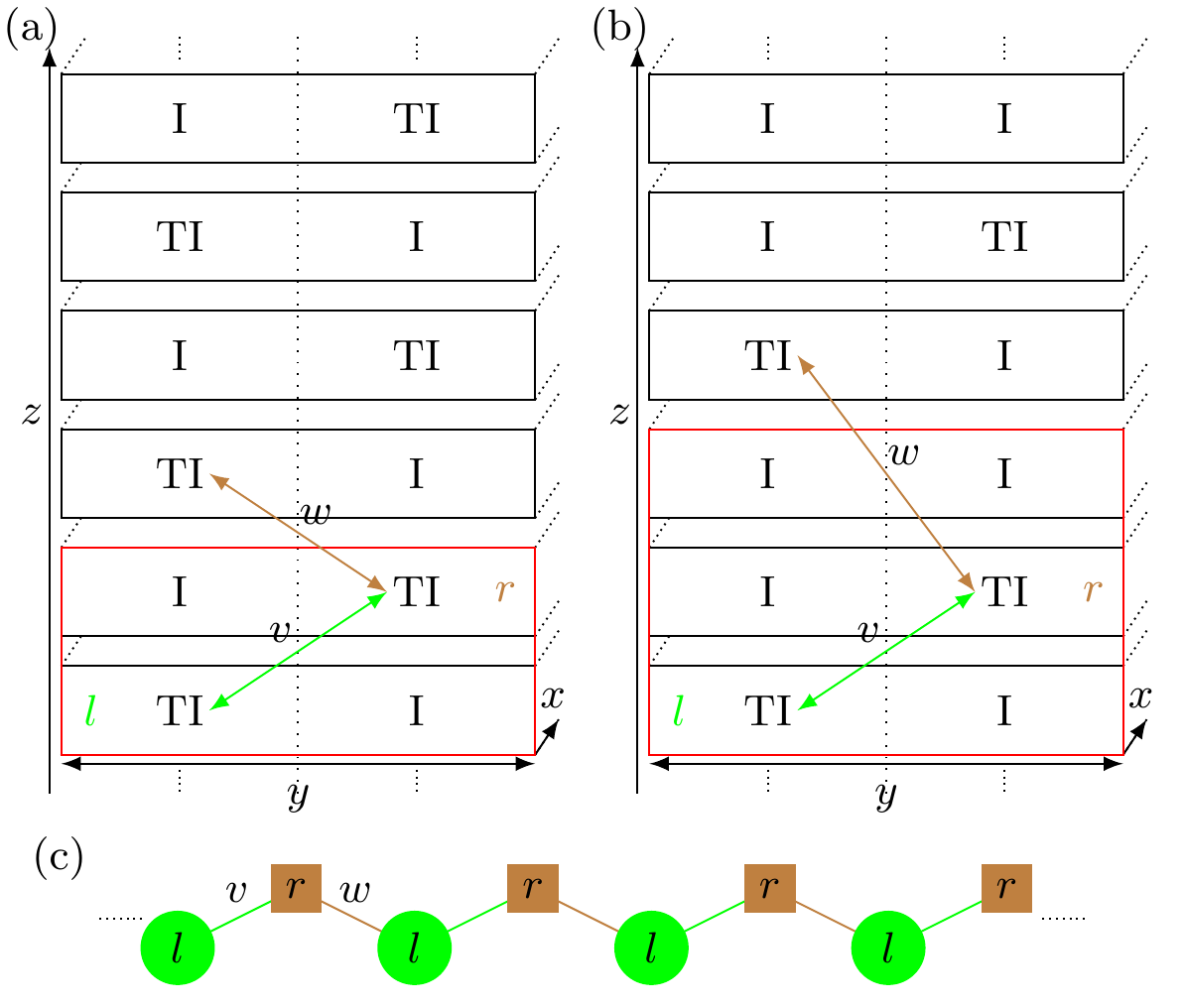}
\caption{
Schematic of the stacking fault construction in the single-spacer WTI (panel a) and in the double-spacer WTI (panel b).
In both cases, the unit cell of the slab is marked in red, topological layers are labeled with TI, and trivial layers with I.
The topological modes of the nontrivial layers on the left ($l$) and right ($r$) side of the planar defect are coupled to each other with a strength $v$ within a unit cell and $w$ between unit cells.
We predict that the stacking fault properties will be governed by the phenomenology of the SSH model (panel c). Thus, the stacking fault will be gapless when $v=w$, as in panel (a), and gapped when $v\neq w$, as in panel (b).
}
\label{fig:WTItoSSH}
\end{figure}

We create a stacking fault in the WTI by using the standard `cut and glue' procedure \cite{Ran2009}.
First, we cut the WTI slab into two halves along its finite direction, ${\bf a_y}$.
One of these halves is then shifted along the stacking direction, ${\bf a_z}$, by a fraction of the unit cell.
For the single-spacer WTI, we shift by one layer, corresponding to half of the unit cell, whereas for the double-spacer WTI the shift is by either one or two layers, corresponding to $\frac13$ or $\frac23$ of a unit cell.
Finally, the two halves are glued back together using the nearest-neighbor hoppings, $t$, that appear both in Eqs.~\eqref{eq:KMmodel} and \eqref{eq:insham}.
Notice that after the stacking fault has been created, each topological layer on one side of the resulting planar defect will have a trivial spacer layer at the same $n_z$ coordinate on the other side.
This construction is shown schematically in Fig.~\ref{fig:WTItoSSH}(a, b) for the single- and double-spacer WTI, respectively.

If the two halves of the system are disconnected from each other (thus, before gluing), each will host two Dirac cones on its surface. 
When these gapless surfaces are reconnected, the two pairs of Dirac cones on either side will couple, producing either a gapped system, or a gapless one.
We predict which of these scenarios will occur based on an analogy to the well-known Su-Schrieffer-Heeger (SSH) model \cite{SSH1968}, sketched in Fig.~\ref{fig:WTItoSSH}(c).
To this end, we imagine the stacking fault to be made up of the edge modes of the individual QSHE layers on either side of the defect, which are coupled to each other with varying strength depending on the distance between them.
The largest of these coupling terms, corresponding to the shortest distances, are labeled as $v$ and $w$ in Fig.~\ref{fig:WTItoSSH}.

\begin{figure}[tb]
\centering
\includegraphics[width=8.6cm]{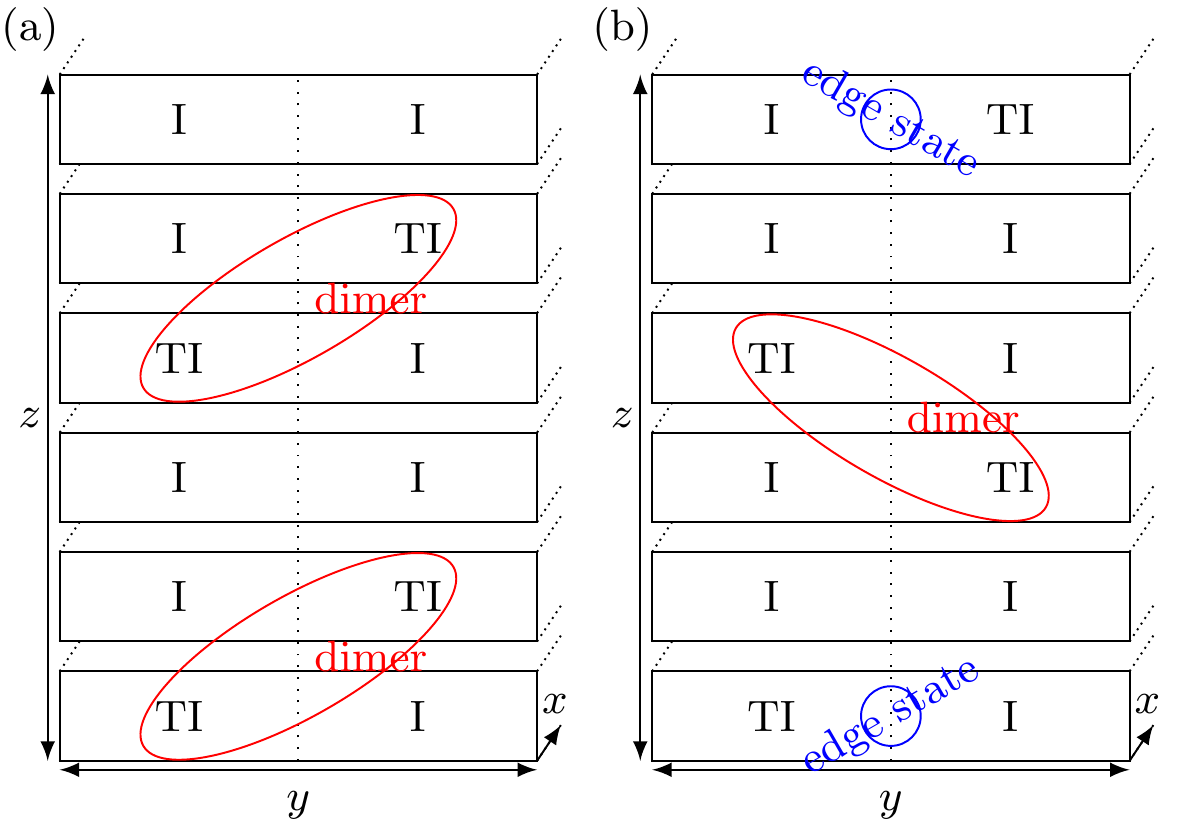}
\caption{
Same as Fig.~\ref{fig:WTItoSSH}(b), but for a slab which is finite in the ${\bf a_z}$ direction, such that it has top and bottom surfaces in addition to the ones in the ${\bf a_y}$ direction.
The red ellipses show the `TI dimers' formed by the unequal coupling strength between edge modes of adjacent topological layers, in analogy to the SSH model.
For a stacking fault as in panel (a), we expect that the planar defect will be gapped and trivial.
For a termination as in panel (b), however, we expect to observe gapless topological modes at the boundaries of the gapped stacking fault (shown in blue).
These are analogous to the edge states of the nontrivial SSH model.
}
\label{fig:stackingfaulttypes}
\end{figure}

Since the QSHE boundary modes are characterized by a $\mathbb{Z}_2$ topological classification, the states in one layer may be gapped out when coupled to those of a nearby layer.
Similar to the sites of the SSH chain, the QSHE edge modes on nearby layers may form dimers when the coupling between them is staggered in the ${\bf a_z}$ direction, as shown in Fig.~\ref{fig:WTItoSSH}(b).
In this case, we expect the stacking fault to be gapped, but its boundaries should host topologically nontrivial states, depending on whether it is terminated with a weak or a strong coupling term (see Fig.~\ref{fig:stackingfaulttypes}).
In contrast, when the boundary modes of adjacent topological layers are equally spaced [$v=w$, as shown in Fig.~\ref{fig:WTItoSSH}(a)], then no dimers are formed, and we expect to observe a gapless stacking fault.
This behavior is again similar to that of the SSH model, which is at a gapless topological phase transition point in the absence of dimerization.

\section{Stacking fault results}
\label{sec:numerics}

\subsection{Single spacer}
\label{subsec:single_spacer}

We begin by considering a stacking fault in the single-spacer WTI model, which is formed between the unit cells at $n_y=19$ and $n_y=20$ in a slab consisting of $L_y=40$ unit cells in the ${\bf a_y}$ direction (in our convention the first unit cell is at $n_y=0$).
Before gluing the two halves of the system back together, the two adjacent surfaces host a total of four Dirac cones, as discussed above.
After including the nearest neighbor hoppings to reconnect the two halves, we find that the two Dirac cones at $(k_x, k_z)=(\pi, 0)$ are still present at the planar defect, while the other two Dirac cones at $(k_x, k_z)=(\pi, \pi)$ gap out.

\begin{figure}[b]
\centering
\includegraphics[width=8.6cm]{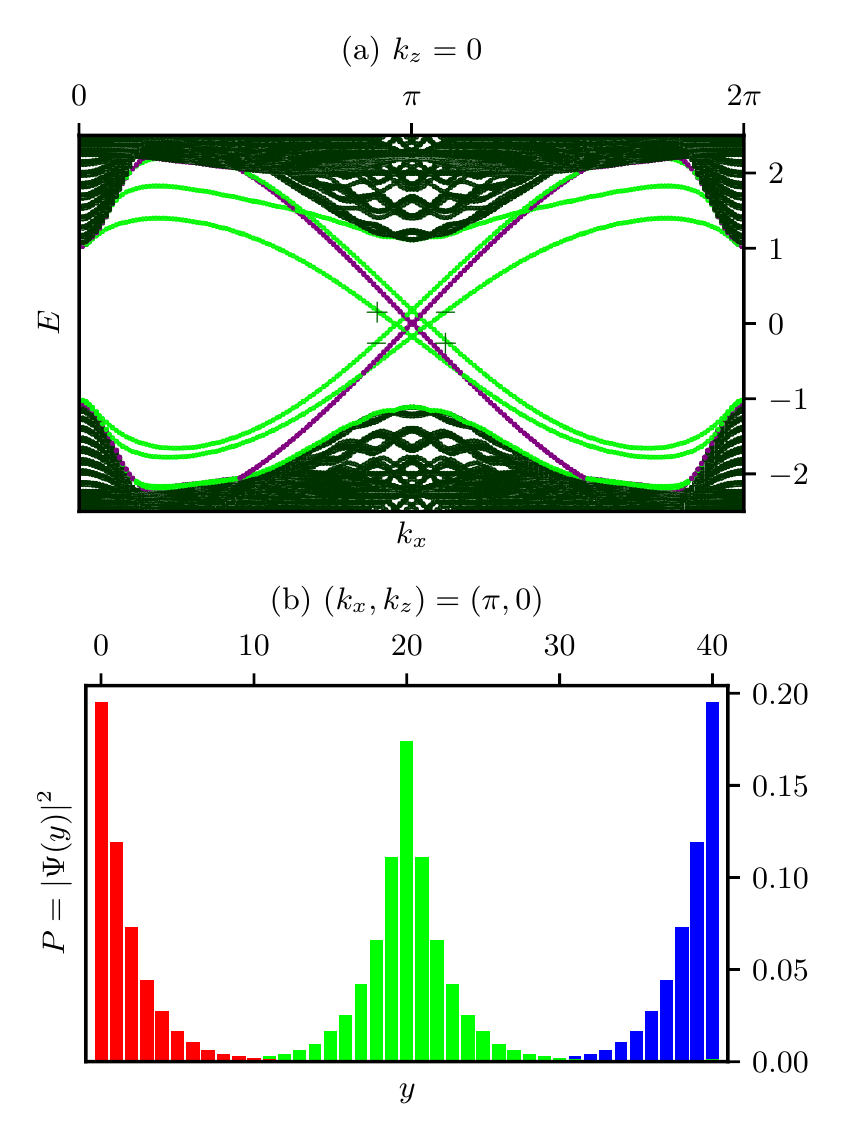}
\caption{
Band structure of the single-spacer WTI with a stacking fault (panel a), computed at $k_z=0$.
Bulk states are shown in dark green, states localized at the surfaces of the slab are shown in purple, whereas modes at the stacking fault are shown in green. 
The $\pm$ indicate the eigenvalues of the mirror symmetry operator $M$, $\pm i$, associated with the topological bands. Since the topological bands have different mirror eigenvalue, they can not gap out unless this symmetry is broken. 
In panel (b) we show the real space probability distribution of the in-gap modes at $(k_x, k_z)=(\pi, 0)$. 
The original WTI surface modes are shown in red and blue, whereas the modes pinned to the stacking fault are shown in green. 
We have used $t_2=1$, $m=2$, $t_{z}=0.5$, $t_{z2}=0.2$ in units of $t$. 
The WTI slab consists of $L_y=40$ unit cells, and the stacking fault is created between the unit cells at $n_y=19$ and $n_y=20$.
}
\label{fig:twolayerstack}
\end{figure}

We show in Fig.~\ref{fig:twolayerstack} the band structure of the system (panel a) at $k_z=0$, as well as the real-space probability distribution of the in-gap modes at the Dirac cone momenta (panel b).
Each surface of the WTI slab still shows gapless modes, as expected, which are colored in red and blue, as in Fig.~\ref{fig:plot_spectra_twolayer}.
In addition, however, there now appear gapless states localized at the stacking fault plane, colored green, which are 2D gapless states embedded in the middle of the gapped, 3D WTI bulk.

The analogy to the 1D SSH model proves to be correct in predicting the gapless nature of the stacking fault in the single-spacer WTI. 
This simple heuristic argument, however, fails to accurately describe the robustness of the stacking fault modes.
In the SSH chain, while the absence of dimerization does produce a gapless phase, the latter is a fine-tuned point in the parameter space, one at which a topological phase transition takes place. 
In contrast, as we show in the following, the 2D gapless phase of the planar defect is robust against perturbations, being protected by mirror symmetry.

Consider a WTI slab in the presence of a stacking fault, infinite along ${\bf a_{x,z}}$ but having a finite width along ${\bf a_y}$, as shown in Fig.~\ref{fig:WTItoSSH}(a). 
The unit cell of the full system (shown in red) is composed of four building blocks: the topological layers on the left and right of the defect, which we label $H_{{\rm KM},l}$ and $H_{{\rm KM},r}$, and the trivial layers on the left and right, labeled $H_{{\rm I},l}$ and $H_{{\rm I},r}$. 
We write the full slab Hamiltonian as 
${\cal H}_{\rm full} = {\bf c}^\dag H_{\rm full} {\bf c}$, 
with ${\bf c}$ a column vector formed by all annihilation operators of the system and $H_{\rm full}$ the Hamiltonian matrix.
In the grading of the four blocks mentioned above this matrix takes the form
\begin{widetext}
\begin{equation}\label{eq:slab_full_H}
 H_{\rm full} = 
 \begin{pmatrix}
  H^\pd_{{\rm KM}, l}+T^\pd_{v2}\sin (k_z) & T^\pd_s & T^\pd_v (1-e^{-ik_z}) & 0   \\
  T^\dag_s & H^\pd_{{\rm I}, r}+T^\pd_{v2}\sin (k_z) & 0 & T^\pd_v (1-e^{-ik_z})   \\
  T^\dag_v (1-e^{ik_z}) & 0 & H^\pd_{{\rm I}, l} +T^\pd_{v2} \sin (k_z)& T^\pd_s   \\
  0 & T^\dag_v (1-e^{ik_z}) & T^\dag_s & H^\pd_{{\rm KM}, r} +T^\pd_{v2}\sin (k_z) \\
 \end{pmatrix}.
\end{equation}
\end{widetext}

Here we have considered that the stacking fault is positioned in the middle of the slab, such that $H_{{\rm KM}, l/r}$ and $H_{{\rm I}, l/r}$ are matrices of equal size, with thicker slabs leading to larger matrices.
The term, $T^\pd_s$ is a Hamiltonian block containing all of the nearest-neighbor hopping terms across the stacking fault, the ones which serve to glue the two halves of the system back together.
$H_{{\rm KM}, l/r}$, $H_{{\rm I}, l/r}$, as well as $T_s$ depend on the momentum $k_x$ but are independent of $k_z$, since they only encode the on-site and hopping terms within the trivial and topological layers, as well as the in-plane hoppings between them.

All of the hoppings along the stacking direction, ${\bf a_z}$, are included in the term $T_v (1-e^{-ik_z})$ appearing in the off-diagonal blocks of Eq.~\eqref{eq:slab_full_H} and in the diagonal term $T^\pd_{v2}\sin (k_z)$.
Part of the off-diagonal term includes the vertical hoppings connecting the trivial and topological layers within the same unit cell, and is thus momentum independent, whereas the second part contains the hoppings connecting layers across the unit cell boundary, hence the factor $e^{-ik_z}$.
These vertical hoppings are given by the inter-layer coupling Hamiltonian of Eq.~\eqref{eq:bihz}, meaning that $T^\pd_v = i t^\pd_z s^z \otimes \mathbb{I}$, with $\mathbb{I}$ an identity matrix of the same size as the number of sites in $H^\pd_{{\rm KM}, l/r}$ and $H^\pd_{{\rm I}, l/r}$. 
It follows that $T^\pd_v = - T^\dag_v$, which explains the relative minus sign.
The diagonal term $T^\pd_{v2}=2t^\pd_{z2}s^y \otimes \mathbb{I}$ contains the next nearest neighbor hoppings connecting the TI (and the trivial) layers to each other. This term obeys $s^zT^\pd_{v2}s^z=-T^\pd_{v_2}$.

Using the form of Eq.~\eqref{eq:slab_full_H} and the properties of $T_v$ and $T_{v2}$, it becomes apparent that the slab Hamiltonian obeys a mirror symmetry. 
The mirror plane is perpendicular to the stacking direction, thus mapping $k_z \to -k_z$, such that
\begin{equation}\label{eq:mirror}
 M^\dag (k_z) H_{\rm full} (k_z) M (k_z) = H_{\rm full} (-k_z).
\end{equation}
The mirror operator, written in the grading of Eq.~\eqref{eq:slab_full_H}, is a block-diagonal matrix composed of four blocks: 
$M = i s^z \otimes {\rm diag} (\mathbb{I}, \mathbb{I}, -\mathbb{I}e^{i k_z}, -\mathbb{I}e^{i k_z})$.
Note that $M$ is a spinful mirror symmetry which commutes with the time-reversal symmetry operator, ${\cal T}=i s^y \otimes \mathbb{I}_{\rm full} {\cal K}$, where $\mathbb{I}_{\rm full}$ is an identity matrix acting on all of the sites of the slab's unit cell and ${\cal K}$ is complex conjugation.

The mirror symmetry protects the Dirac cones at $k_z=0$ but not the ones at $k_z=\pi$.
On the mirror-invariant plane of the Brillouin zone, $k_z=0$, the trivial and topological layers are effectively decoupled in the vertical direction.
The states of $H_{{\rm KM}, l}$ and $H_{{\rm I}, r}$ are associated to a mirror eigenvalue $+i$, whereas those of $H_{{\rm KM}, r}$, $H_{{\rm I}, l}$ are orthogonal to them, having a mirror eigenvalue $-i$ (as shown in Fig.~\ref{fig:plot_spectra_twolayer}).
Mirror symmetry thus prevents the topological modes of the QSHE layers on either side of the stacking fault from coupling to each other, guaranteeing the formation of gapless phase at the planar defect.
The latter forms an embedded, 2D topological semimetal protected by mirror symmetry. 
In contrast, at $k_z=\pi$, the QSHE modes on the left and right of the stacking fault have the same mirror eigenvalues, and are therefore not protected.

\subsection{Double spacer}
\label{subsec:double_spacer}

We now consider a stacking fault in the two-spacer WTI.
In this case we shift half of the system in the ${\bf a_z}$ direction by either $\frac{1}{3}$ or $\frac{2}{3}$ of the unit cell. When the system has an infinite number of layers, these two scenarios differ only in the choice of the unit cell. 
In both cases, the corresponding infinite SSH chain is in a insulating phase and the spectrum of the system should be gapped. 

We have created, as before, two identical copies of the system shifted with respect to each other in the ${\bf a_z}$ by $\frac{1}{3}$ of the unit cell. In this way every TI layer on one side of the stacking fault has a trivial layer on the opposite side.
The two sides are glued together, as before, using the nearest-neighbor hopping $t$. 

In this case, the helical modes at the stacking fault develop a gap at both the Dirac points at $k_z=0,\pi$, and the system has no mirror symmetry. 
We obtain an embedded SSH-like system with $v\neq w$ at the defect, as we expected. 

\begin{figure}[tb]
\centering
\includegraphics[width=8.6cm]{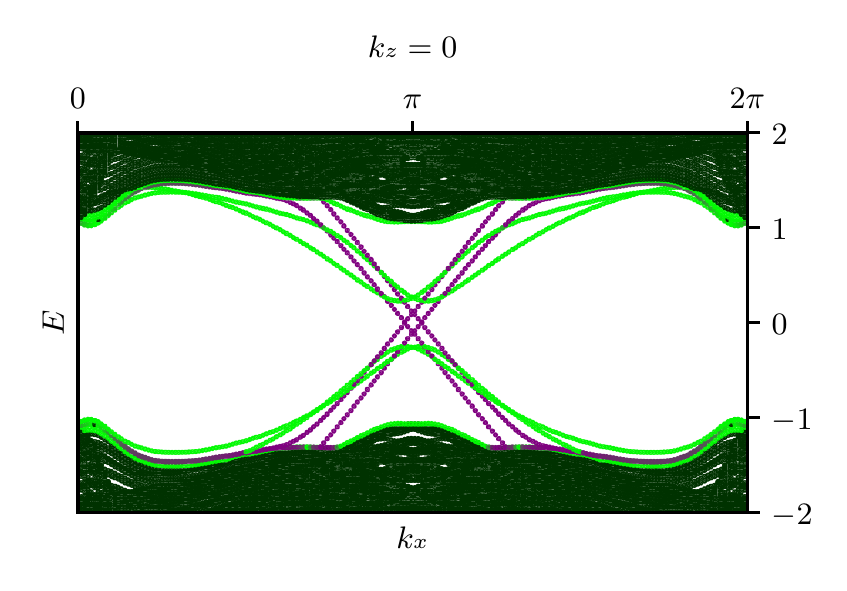}
\caption{
Band structure of the double-spacer WTI with a stacking fault, computed at $k_z=0$.
Bulk states are shown in dark green, states localized at the surfaces of the slab are shown in purple, whereas modes at the stacking fault are shown in green. 
In this case there is no mirror symmetry protecting the states at the stacking fault and they gap out. 
The states at the external surface are still present because they are protected by the time-reversal and translation symmetry along the stacking direction.
We have used the inter-layer coupling Hamiltonian in Eq.~(\ref{eq:threehzgap}) with $t_2=1$, $m=2$, $t_{z}=0.8$, $t_{z2}=0.2$, $t_{z3}=0.2$ in units of $t$. 
The WTI slab consists of $L_y=40$ unit cells, and the stacking fault is created between the unit cells at $n_y=19$ and $n_y=20$.
}
\label{fig:threelayerstack}
\end{figure}

To distinguish between the topological and trivial phase of the SSH model, we study a system with a finite number of layers to detect the boundary states at the beginning and at the end of the stacking fault. 
In a system with a finite number of layers we will have more bands near the the Fermi level $E=0$. 
In order to distinguish the bands associated with the topological states at the boundaries of the stacking fault, is convenient to make the gap larger. 
We achieve this by replacing the Hamiltonian in Eq.~\eqref{eq:threehz} with:
\begin{equation}
\begin{split}
\widetilde{H}_z & = \sum_{{\bf r}, j, \alpha, \beta} it^\pd_z c^\dagger_{{\bf r}, j,\alpha} c^\pd_{{\bf r}+{\bf a_z}, j,\beta} s^z_{\alpha, \beta}\lambda_j \\
& + it^\pd_{z2} c^\dagger_{{\bf r}, j,\alpha} c^\pd_{{\bf r}+2{\bf a_z}, j,\beta} s^z_{\alpha, \beta}\lambda_j \\
&+it^\pd_{z3} c^\dagger_{{\bf r}, j,\alpha} c^\pd_{{\bf r}+3{\bf a_z}, j,\beta} (s^y_{\alpha, \beta}+s^z_{\alpha, \beta}\lambda_j)/\sqrt{2} +{\rm h. c.}
\end{split}
\label{eq:threehzgap}
\end{equation}
where $\lambda_j=\pm 1$ for sites that belong to the A or B sublattice respectively. 
The new term, $s^z_{\alpha, \beta}\lambda_j$, couples equal-spin components of the surface states. 
Components with the same spin of helical modes at opposite sides of the defect will have opposite velocity, so this new term will produce a bigger gap at the stacking fault, see Fig.~\ref{fig:threelayerstack}.

\subsection{Topological edge states at the stacking fault}
\label{subsec:stacking_fault_edges}

\begin{figure}[tb]
\centering
\includegraphics[width=8.6cm]{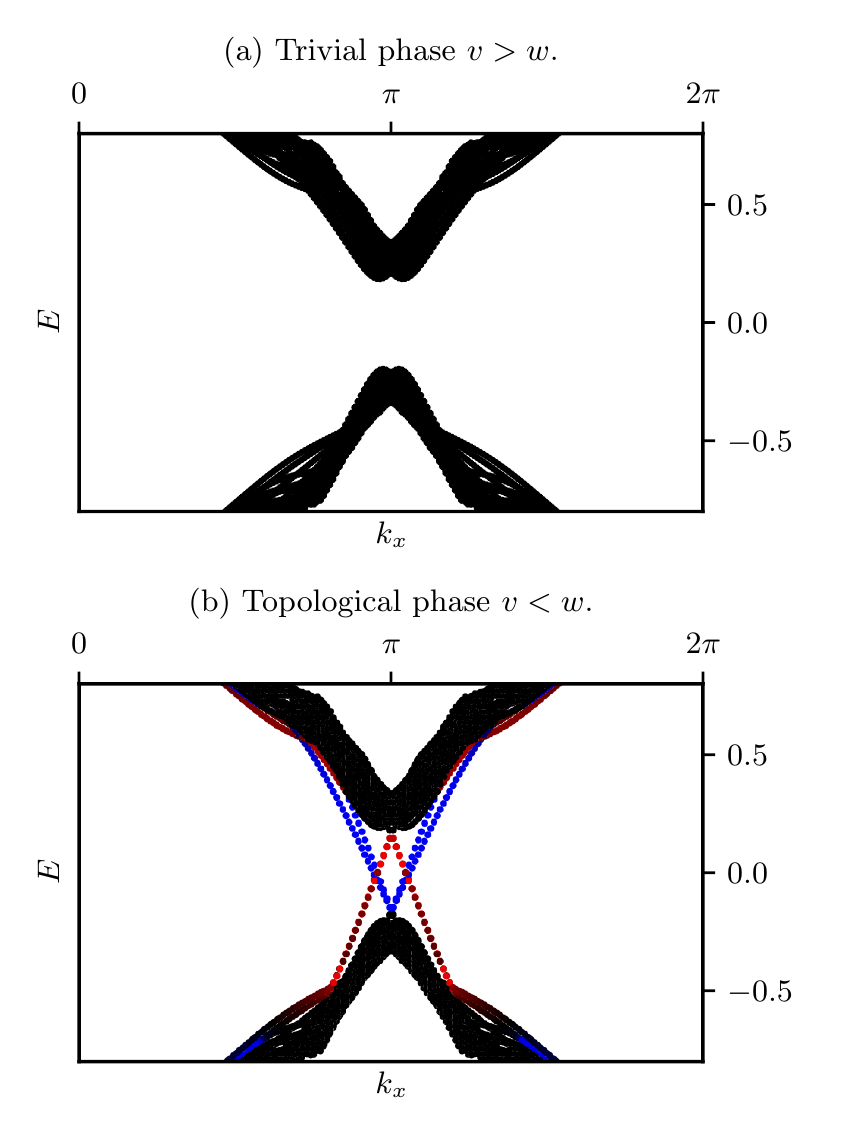}
\caption{
Spectra of the WTI with two spacers per unit cell with a finite number of layers. 
Both spectra are obtained adding periodic boundary conditions in the ${\bf a_y}$ direction. 
In (a) the defect begins and ends with dimers of TI layers and the corresponding SSH chain is in the trivial phase with $v>w$. 
In (b) the SSH chain is in the topological phase $v<w$ and the gapless states at the boundaries of the two defects are clearly visible. 
The red (blue) bands are associated with states localized at the top (bottom) side of the stacking fault at $n_y=19$ and the bottom (top) side of the one at $n_y=39$, see Fig.~\ref{fig:2d_probability}.
Both plots are obtained for $t_2=1$, $m=2$, $t_{z1}=0.8$, $t_{z2}=0.2$ $t_{z3}=0.2$ in units of $t$. The WTI has $L_y=40$ and $L_z=30$.
}
\label{fig:topo-trivial-stack}
\end{figure}

In order to check the presence of topological states at the defect in the tight-binding model, we build a system with a finite height, $L_z$. 
To better visualize the stacking fault modes, we add periodic boundary conditions in the ${\bf a_y}$ direction, such that the system contains two stacking faults. 
In this way we eliminate the surface modes and obtain only the spectrum of the two defects.

The SSH chain associated with the WTI with two trivial spacers will be in the trivial ($v>w$) or in the topological ($v<w$) phase depending on the boundary conditions at the stacking fault. 
If the defect begins and ends with a couple of dimerized TI layers, we do not find any gapless states (see Fig.~\ref{fig:topo-trivial-stack}), similar to the infinite case.

\begin{figure}[tb]
\centering
\includegraphics[width=8.6cm]{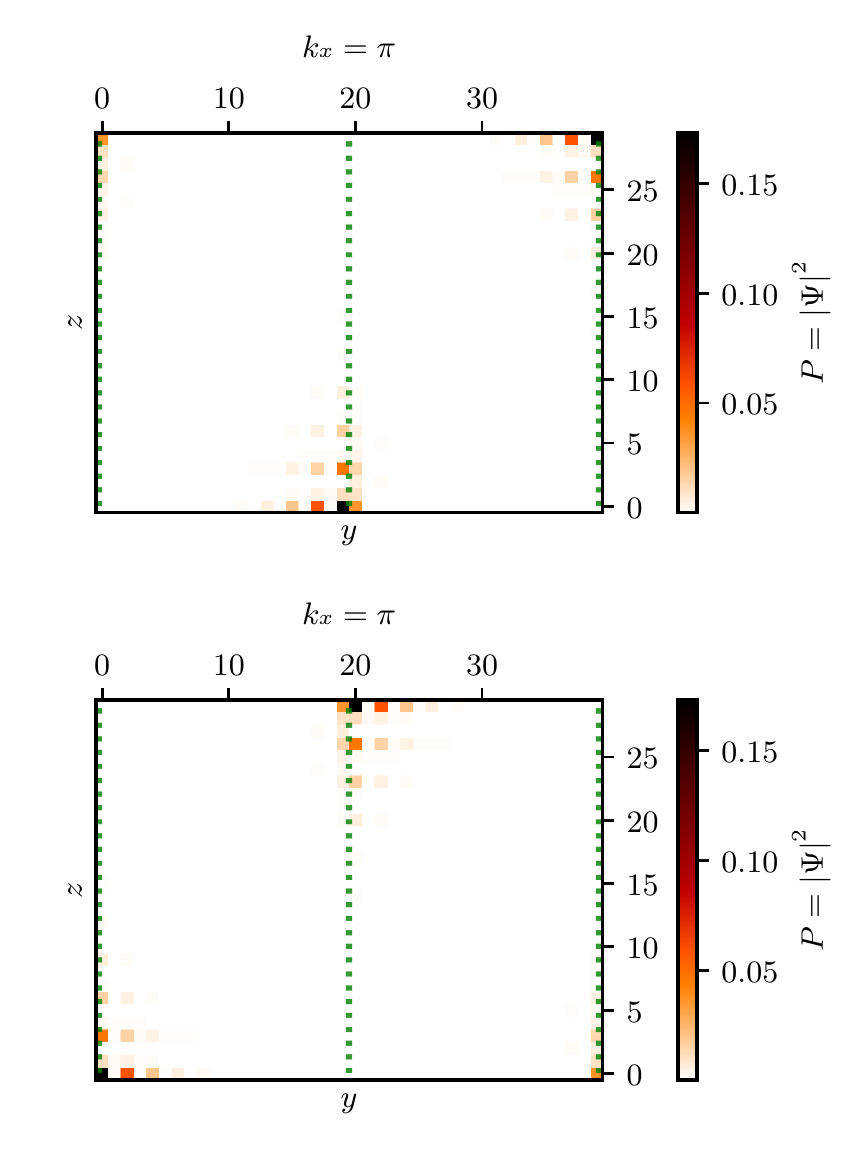}
\caption{
Probability distribution of the gapless states $P(x,z)=|\braket{x,z|\Psi}|^2$ as a function of the position $(y,z)$ in the unit cell. 
The system has $L_y=40$ and $L_z=30$, and two stacking faults are created, one between the unit cells at $n_y=19$ and $n_y=20$ and a second one between $n_y=39$ and $n_y=0$, adding periodic boundary conditions in the $y$ direction. The stacking faults are highlighted in green in the plot.
The topological states are localized at the decoupled TI layers at the edges of the two stacking faults, as expected. 
The states on different decoupled edge TIs have different energies. 
The top figure corresponds to blue bands in Fig.~\ref{fig:topo-trivial-stack} and the bottom one corresponds to the red ones.
Both plots are obtained for $t_2=1$, $m=2$, $t_{z1}=0.8$, $t_{z2}=0.2$ $t_{z3}=0.2$ in units of $t$. 
}
\label{fig:2d_probability}
\end{figure}

For a system with two weakly-coupled TI layers at the boundary of the defect, the corresponding SSH chain is in the topological phase. 
In this case, we find topological gapless bands located at the boundaries of the two stacking faults, see Fig.~\ref{fig:topo-trivial-stack}.
We confirm their location by plotting the real-space probability distribution, $P(x,z)=|\braket{x,z|\Psi}|^2$, as a function of the position $(y,z)$ in the unit cell in Fig.~\ref{fig:2d_probability}.

\section{Conclusions}
\label{sec:conc}

In order to explore the effect of stacking faults in weak topological insulators like $\textrm{Bi}_2\textrm{TeI}$ and $\textrm{Bi}_{14}\textrm{Rh}_3\textrm{I}_9$, we have used a tight binding model composed of stacked honeycomb QSHE layers.
We have then used a heuristic argument to draw an analogy between the physics of the topological states at the defect and an SSH chain. 

The stacking fault in a WTI with one trivial spacer per unit cell behaves like an SSH at the gap closing point. Thus, it forms an embedded topological semimetal, which we have shown to be protected by mirror symmetry.
In contrast, a WTI with two insulating spacers per unit cell behaves like an SSH chain in the topological or trivial phase, depending whether or not it has weakly-coupled TI layers at the edges of the stacking fault.
In the latter case, an embedded topological insulator is formed: the stacking fault hosts 1D gapless modes at its boundaries.

This means that it is in principle possible to obtain embedded topological insulators or semimetals inside the bulk of a WTI in the presence of a stacking fault. 
In this case the defect is not destroying the topological properties of the system, but is a source of new, robust states.

Our toy model is only qualitative in nature. 
A quantitative prediction of the real properties of stacking faults in $\textrm{Bi}_2\textrm{TeI}$ and $\textrm{Bi}_{14}\textrm{Rh}_3\textrm{I}_9$ would require more accurate numerical methods, such as density functional theory.
Nevertheless, the analogy to the SSH model may still be used to guide this endeavor.
For instance, while stacking faults obtained by a fractional translation have not been reported to date in these materials, it has been shown that Bi$_2$TeI forms inverted twin domains \cite{Zeugner2017, Zeugner2018}.
Since its unit cell is composed of one QSHE layer and two spacers, we conjecture that these stacking faults will host protected gapless modes at their boundaries whenever they are terminated by weakly-coupled TI layers.

\section{Acknowledgments}

We thank Ulrike Nitzsche for technical assistance. 
We acknowledge financial support from the DFG through the W\"urzburg-Dresden Cluster of Excellence on Complexity and Topology in Quantum Matter -- \textit{ct.qmat} (EXC 2147, project-id 39085490).

\bibliography{ref}

\end{document}